\documentclass{PoS}

\usepackage{amsmath}
\newcommand{\sgn}{\mbox{sgn}}
\newcommand{\be}{\begin{equation}}
\newcommand{\ee}{\end{equation}}
\newcommand{\bea}{\begin{eqnarray}}
\newcommand{\eea}{\end{eqnarray}}
\newcommand{\bt}{\begin{tabbing}}
\newcommand{\et}{\end{tabbing}}
\newcommand{\bi}{\begin{itemize}}
\newcommand{\ei}{\end{itemize}}
\newcommand{\ben}{\begin{enumerate}}
\newcommand{\een}{\end{enumerate}}

\title{%
   \begin{picture}(0,0)(0,0)%
   \put(355,75){\makebox(0,0)[l]{\textnormal{\normalsize KEK-CP-180}}}%
   \put(355,60){\makebox(0,0)[l]{\textnormal{\normalsize RIKEN-TH-77}}}%
   \put(355,45){\makebox(0,0)[l]{\textnormal{\normalsize UTHEP-527}}}%
   \end{picture}%
   JLQCD's dynamical overlap project
}

\ShortTitle{JLQCD's dynamical overlap project}

\author{
   JLQCD Collaboration: 
   \speaker{T.~Kaneko}$^{a,b}$\thanks{E-mail: takashi.kaneko@kek.jp},
   S.~Aoki$^c$, 
   H.~Fukaya$^d$,
   S.~Hashimoto$^{a,b}$, 
   K-I.~Ishikawa$^e$, 
   K.~Kanaya$^c$, 
   H.~Matsufuru$^a$, 
   M.~Okamoto$^a$, 
   M.~Okawa$^e$, 
   T.~Onogi$^g$, 
   A.~Ukawa$^{f,c}$, 
   N.~Yamada$^{a,b}$,
   T.~Yoshi\'e$^{f,c}$
   \\
   \\
   \llap{$^a$}
   High Energy Accelerator Research Organization (KEK),
   Ibaraki 305-0801, Japan 
   \\
   \llap{$^b$}
   School of High Energy Accelerator Science,
   The Graduate University for Advanced Studies (Sokendai),
   Ibaraki 305-0801, Japan
   \\
   \llap{$^c$}
   Graduate School of Pure and Applied Sciences,
   University of Tsukuba, Ibaraki 305-8571, Japan
   \\
   \llap{$^d$}
   Theoretical Physics Laboratory, RIKEN, Saitama 351-0198, Japan
   \\
   \llap{$^e$}
   Department of Physics, Hiroshima University, Hiroshima 739-8526, Japan
   \\
   \llap{$^f$}
   Center for Computational Sciences,
   University of Tsukuba, Ibaraki 305-8577, Japan
   \\
   \llap{$^g$}
   Yukawa Institute for Theoretical Physics, Kyoto University,
   Kyoto 606-8502, Japan
}

\abstract{
We present an overview of our project of large-scale simulations with
dynamical overlap fermions. The first production run in two-flavor QCD 
is on-going using the Iwasaki gauge action 
on a $16^3 \times 32$ lattice at the lattice spacing of 0.12~fm
with six sea quark masses down to $m_{s,\rm phys}/6$, where $m_{s,\rm phys}$
is the physical strange quark mass.
We briefly introduce our choice of the lattice action and simulation algorithm,
and describe the present status of the production run.
Preliminary results on the light meson masses and the static quark potential 
are also reported.
}

\FullConference{XXIVth International Symposium on Lattice Field Theory\\

                July 23-28, 2006\\

                Tucson, Arizona, USA}

\begin{document}

%// introduction --------------------------------------------------------------

\section{Introduction}

The JLQCD Collaboration has pursued detailed studies of 
non-perturbative properties of QCD 
through numerical simulations on the supercomputer system at KEK.
Our previous unquenched simulations 
\cite{Spectrum:Nf2QCD:JLQCD,Spectrum:Nf3QCD:CP-PACS+JLQCD}
however were restricted to rather heavy sea quark masses 
$\gtrsim m_{s, \rm phys}/2$,
which make the chiral extrapolation difficult to control.
The use of Wilson-type quarks also limits the applicability of our simulation
to quantities, such as $B_K$, for which 
the explicit chiral symmetry breaking causes complicated operator mixing.

This year, the supercomputer system at KEK was replaced by a multiple system
of Hitachi SR11000 and IBM Blue Gene/L with a total peak speed of about 
60 TFLOPS.
On this new system,
we carry out large-scale simulations including dynamical overlap 
fermions with sea quark masses down to $m_{s, \rm phys}/4$ or smaller
on reasonably fine ($a\!\lesssim\!0.125$~fm) and 
large lattices ($L\!\gtrsim\!2$~fm).
In this article, 
we present an overview of our first production simulation in two-flavor QCD.

%// simulation method ---------------------------------------------------------

\section{Simulation method}

%// choice of lattice action ------------------------------------------

%// gauge action

Our numerical simulations are carried out using the 
overlap quark action % \cite{overlap}
with the standard Wilson kernel $H_W$
\bea
   D(m_{\rm sea})
   & = &
   \left(m_0+\frac{m_{\rm sea}}{2}\right)
  +\left(m_0-\frac{m_{\rm sea}}{2}\right)\,
      \gamma_5\, \sgn\left[ H_W(-m_0) \right],
   \label{eqn:overlap}
\eea
where we set $m_0\!=\!1.6$.
A major problem with this kernel is the appearance of 
(near-)zero modes of $H_{\rm W}$ in simulations 
on relatively coarse lattices.
This possibly spoils the locality of $D$ 
and also makes simulations costly.
From our preparatory study in quenched QCD \cite{lat06:JLQCD:yamada}, 
we adopt the Iwasaki gauge action, 
with which the near-zero modes are suppressed compared to the standard 
plaquette glue and the localization range of $D$ is 
sufficiently small.
% the mobility edge \cite{mobility_edge} is not so small
% ($\approx$ 600~MeV).

%// extra-Wilson
The use of the improved gauge action does not completely rule 
out the appearance of exact zero modes.
In order to avoid the time consuming reflection/refraction procedure
\cite{reflect},
we introduce two-flavors of unphysical extra-Wilson fermions with 
a large negative mass $-m_0$ \cite{extra-Wilson:else,extra-Wilson:JLQCD}
and additional twisted mass ghosts
\cite{extra-Wilson:JLQCD} which produce the Boltzmann weight
\bea
   \frac{\det\left[ H_{\rm W}(-m_0)^2 \right]}
        {\det\left[ H_{\rm W}(-m_0)^2 + \mu^2 \right]}.
   \label{eqn:extra-Wilson}
\eea
The factor in the numerator suppresses the zero modes during continuous 
updating of gauge fields, such as the hybrid Monte Carlo (HMC) algorithm.
The denominator with appropriately chosen twisted mass $\mu$ cancels
possibly large effects from high modes of $H_W$ in the numerator.
We note that these unphysical extra-fields have masses of cutoff order, 
and hence their effects vanish in the continuum limit.
The suppression of zero modes in quenched and two-flavor QCD
is demonstrated in Refs.\cite{extra-Wilson:JLQCD,lat06:JLQCD:hashimoto}.

While this procedure fixes the net topological charge $Q$ 
during the HMC update,
it does not forbid local topological fluctuations.
It is expected that, in the infinite volume limit, 
properties of hadrons with their size of $O(1\,{\rm fm})$ 
are insensitive to the global topological charge of gauge configurations,
{\it i.e.} the effect due to the fixed topology is a finite size effect (FSE).
We refer to Ref.\cite{lat06:JLQCD:hashimoto} for more detailed discussions.
We also note that our lattice action to fix the topology 
provides a convenient framework for studies in the $\epsilon$-regime 
\cite{lat06:JLQCD:fukaya}.

%// algorithm ---------------------------------------------------------

%// mult, solver

In our simulation program,
we implement multiplications of $D$ employing 
the low mode preconditioning: 
namely, we introduce a threshold $\lambda_{\rm th}$ in 
the spectrum of $H_{\rm W}$, and % eigenvectors with their 
eigenvalues smaller than $\lambda_{\rm th}$ are projected out
in the evaluation of $\sgn[ H_W ]$.
For higher modes,
we use the rational approximation with the Zolotarev coefficients
and the multi-shift CG algorithm for the inner-loop \cite{multi_shift_solver}.
For multiplications of $D^{-1}$,
we employ a nested solver 
with the relaxed CG \cite{relaxed_solver} for the outer-loop.

%// HMC

\begin{table}[b]
   \begin{center}
   \begin{tabular}{l|llll|lll|ll}
   \hline
   $m_{\rm sea}$  & $m^\prime$
                  & $N_{\rm MD}^{\rm (PF2)}$
                  & $R_{\rm MD}^{\rm (PF)}$
                  & $R_{\rm MD}^{\rm (G)}$
                  & HMC traj.
                  & $P_{\rm HMC}$
                  & time[min]
                  & \#conf$_{\rm pot}$
                  & \#conf$_{\rm had}$
   \\ \hline
   0.015  & 0.2 & 9 & 4 & 5  & 2150 & 0.89 & 6.1 & 151 & 173
   \\
   0.025  & 0.2 & 8 & 4 & 5  & 4320 & 0.90 & 4.7 & 389 & 410
   \\
   0.035  & 0.4 & 6 & 5 & 6  & 4150 & 0.74 & 3.0 & 411 & 403
   \\
   0.050  & 0.4 & 6 & 5 & 6  & 3500 & 0.79 & 2.6 & 310 & 310
   \\
   0.070  & 0.4 & 5 & 5 & 6  & 3500 & 0.81 & 2.1 & 307 & 243
   \\
   0.100  & 0.4 & 5 & 5 & 6  & 3590 & 0.85 & 2.0 & 301 & 319
   \\ \hline
   \end{tabular}
   \caption{
      Simulation parameters for configuration generation.
      Statistics for the measurements of static potential 
      (\#conf$_{\rm pot}$) and hadron correlators (\#conf$_{\rm had}$)
      are also listed.
   }
   \vspace{0mm}
   \label{tbl:sim_param}
\end{center}
\end{table}

In our implementation of HMC,
we employ a combination of the Hasenbusch preconditioning 
\cite{mass_precond} and the multiple time scale discretization
of the molecular dynamics (MD) \cite{mtsMD},
which has been shown to be effective in simulations with 
Wilson-type fermions \cite{mass_precond+mtsMD}.
The determinant factor for overlap quarks is written as 
\bea
   \det\left[D(m_{\rm sea})^2\right]
   & = & 
   \det\left[D(m^\prime)^2\right]
   \det\left[\frac{D(m_{\rm sea})^2}
                  {D(m^\prime)^2}\right],
   \label{eqn:mass_precond}
\eea
where $m^\prime$ is the mass of the Hasenbusch preconditioner.
Two pseudo-fermions, which we call PF1 and PF2 in the following, 
are introduced for the first and second factors in the r.h.s. 
Then, PF2 is updated $N_{\rm MD}^{({\rm PF2})}$ times 
per unit trajectory length. The updates of PF1 and gauge fields 
are more frequent by factors of $R_{\rm MD}^{({\rm PF})}$ and 
$R_{\rm MD}^{({\rm PF})}\,R_{\rm MD}^{({\rm G})}$.
% , where $R_{\rm MD}^{({\rm PF})}$ and $R_{\rm MD}^{({\rm g})}$ 
% are a positive integer.
%
Since we suppress zero modes by the extra-Wilson fermion,
we switch off the reflection/refraction procedure in our 
MD evolution.

%// assembler code 

In addition to the above algorithmic techniques,
we also use an assembler code developed by IBM for multiplications
of $D_W$ on Blue Gene/L.
This code makes the best use of the double FPU instructions,
which process double precision complex number arithmetic effectively
using two sets of floating-point registers.
This assembler code is a factor of 3 faster than our naive Fortran code.
Further details of our simulation algorithm are presented 
in Ref.~\cite{lat06:JLQCD:matsufuru}.

%// production run ------------------------------------------------------------

\section{Production run}

%// simulations parameters --------------------------------------------

Our first production run with two-flavors of degenerate up and down 
quarks is being carried out with the above mentioned lattice action.
The twisted mass $\mu\!=\!0.2$ is chosen from a preparatory study 
in quenched QCD \cite{extra-Wilson:JLQCD}.
We simulate $\beta\!=\!2.30$, which is expected to correspond to the 
lattice spacing about 0.125~fm, on a $16^3 \times 32$ lattice.
Six sea quark masses listed in Table~\ref{tbl:sim_param}
are taken to explore a range from $m_{s, \rm phys}$ down to 
$m_{s, \rm phys}/6$.
All results in this article are obtained in the trivial topological sector 
$Q\!=\!0$.

\FIGURE{
   \centering
   % \begin{figure}[h!]
   % \begin{center}
   \includegraphics[angle=0,width=0.45\linewidth,clip]{F_b230_m0015.eps}
   \vspace{-3mm}
   \caption{
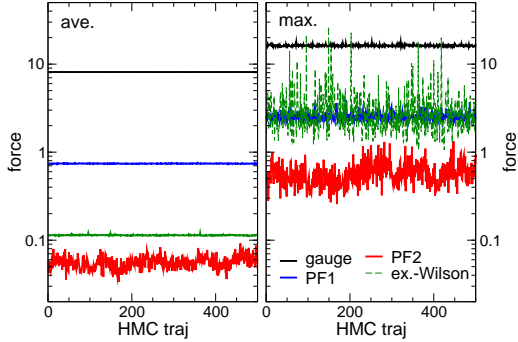
      Average (left panel) and maximum value (right panel) of MD force 
      at each trajectory. Data at $m_{\rm sea}\!=\!0.015$ are plotted.
   }
   \label{fig:sim:force}
   % \end{center}
   % \end{figure}
}

In all simulations, we fix $\lambda_{\rm th}\!=\!0.045$,
with which several eigenvectors are projected out in 
the calculation of $\sgn[H_{\rm W}]$.
The number of poles in the Zolotarev approximation is set to 10
leading to an accuracy of 
$|\sgn[H_{\rm W}]^2\!-\!1| \simeq 10^{-(7\mbox{\,-\,}8)}$ 
throughout the simulations.
The stopping condition of the overlap solver is chosen so that 
the reversibility in the gauge field and the Hamiltonian 
is satisfied to a level comparable to our previous simulations.

We are generating gauge configurations on Blue Gene/L and 
store them on disks every 10 trajectories.
The HMC trajectory length is fixed to 0.5.
Hadron correlators and the static quark potential are measured 
on both of SR11000 and Blue Gene/L using the stored configurations.
Current statistics % together with parameters for the MD evolution 
are summarized in Table~\ref{tbl:sim_param}.

Figure~\ref{fig:sim:force} shows 
an example of the time history of the MD force from gauge fields, 
PF1 and PF2.
With our choice of $m^\prime$, these forces are well separated 
from each other. 
This situation suitable for the application of the multiple time scale 
discretization is confirmed also at other sea quark masses.
In the same figure, we also plot the force from the extra-fields 
in Eq.~(\ref{eqn:extra-Wilson}).
Its maximum value becomes as large as that for the gauge field
probably due to small eigenvalues of $H_{\rm W}$.
Therefore, we update the extra-fermions in the inner most MD step.

\FIGURE{
   \centering
   % \begin{figure}[h!]
   % \begin{center}
   \includegraphics[angle=0,width=0.45\linewidth,clip]{tau_vs_mud.eps}
   \vspace{-3mm}
   \caption{
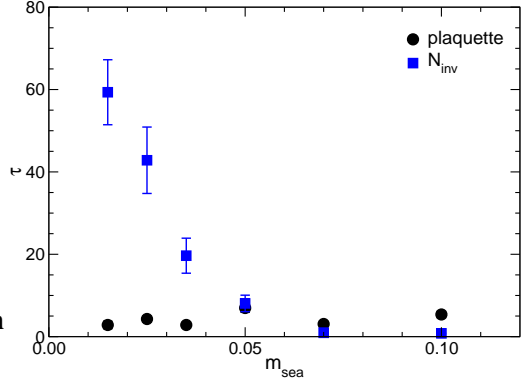
      Autocorrelation time for plaquette and iteration count 
      of the overlap solver $N_{\rm inv}$.
   }
   \label{fig:sim:tau}
   % \end{center}
   % \end{figure}
}

As summarized in Table~\ref{tbl:sim_param},
we achieve the acceptance rate $P_{\rm HMC}$ of 74~\% or higher
with our choice of the MD discretization parameters.
The CPU time per 1 trajectory on the whole machine of Blue Gene/L
(10 racks, 57.3~TFLOPS) is also listed in the table.
It takes about one month to accumulate 2000 trajectories at six sea quark 
masses.
We have recently switched to a five dimensional solver 
proposed in Ref.\cite{5D_solver},
which reduces the CPU time 
by roughly a factor of 2 for our simulation parameters.

%// autocorrleation ---------------------------------------------------

In Fig.~\ref{fig:sim:tau},
we plot the sea quark mass dependence of the autocorrelation time for 
two quantities: $\tau_{\rm plaq}$ for the plaquette, and 
$\tau_{\rm inv}$ for the iteration count of the overlap solver $N_{\rm inv}$.
We find that 
$\tau_{\rm plaq}$ is small and has a mild mass dependence,
probably because it is a local quantity.
On the other hand, $N_{\rm inv}$ is expected to be sensitive to 
the low modes of $D$.
In fact, $\tau_{\rm inv}$ increases rapidly towards the chiral limit.

% We find that $\tau_{\rm inv}$ is around 60 trajectories 
We find that $\tau_{\rm inv} \approx 60$ at our lightest sea quark mass.
This implies that statistics of $O(10,000)$ trajectories 
are needed for precise calculation of hadronic observables
at such light quark masses.
In the following, we present preliminary results with limited statistics 
listed in Table~\ref{tbl:sim_param}.

%// static quark potential ----------------------------------------------------

\section{Static quark potential}

%// potential to lattice spacing --------------------------------------

We calculate the static quark potential $V(r)$ from the smeared Wilson loops.
% with the smearing in Ref.\cite{smeaing}. 
The Sommer scale $r_0$ \cite{r0} is extracted from a parametrization
$V(r)\!=\!V_0 - \alpha/r + \sigma\,r$.

\begin{figure}[t]
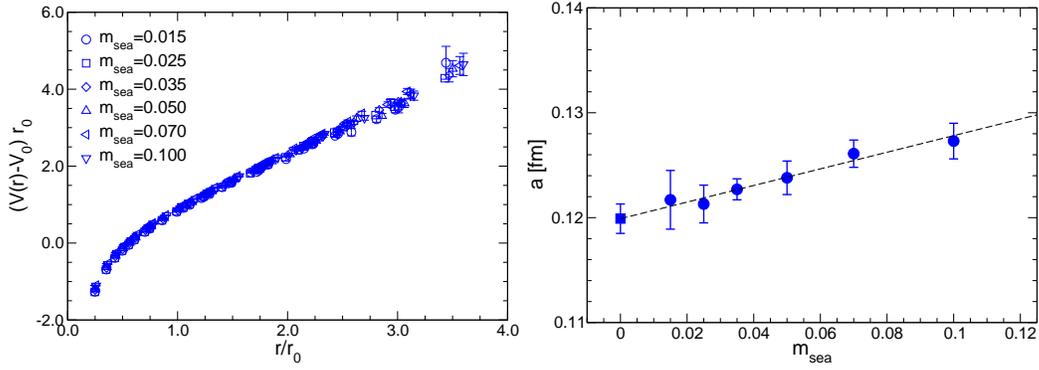

\begin{center}
   \includegraphics[angle=0,width=0.45\linewidth,clip]{VvsR_b230_rescale.eps}
   \includegraphics[angle=0,width=0.45\linewidth,clip]{a_vs_m_b230.eps}
   \vspace{-3mm}
   \caption{
      Plot of $(V(r)\!-\!V_0)\,r_0$ as a function of $r/r_0$ (left figure), 
      and chiral extrapolation of lattice spacing determined from $r_0$ 
      (right figure).
   }
   \vspace{-5mm}  
   \label{fig:pot}
\end{center}
\end{figure}

In Fig.~\ref{fig:pot}, we plot $V(r)$ rescaled by $r_0$
at all sea quark masses. 
While we simulate a wide range of the sea quark mass, 
all data form a universal curve.
We note that our statistics are not sufficiently high to observe 
a clear dependence of $\alpha$ on the sea quark mass.
The chiral extrapolation of the lattice spacing determined from $r_0$ 
is shown in the same figure.
From a simple linear fit and an input $r_0\!=\!0.49$~fm,
we obtain $a\!=\!0.1199(14)$~fm in the chiral limit,
which is close to our target value 0.125~fm.

\begin{figure}[b]
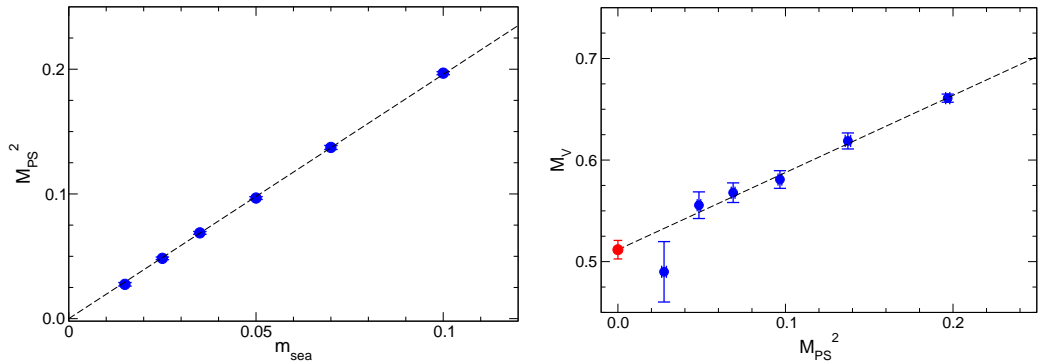

\begin{center}
   \includegraphics[angle=0,width=0.45\linewidth,clip]{mPS2_vs_mud.eps}
   \hspace{2mm}
   \includegraphics[angle=0,width=0.43\linewidth,clip]{mV_vs_mPS2.eps}
   \vspace{-3mm}
   \caption{
      Chiral extrapolation of pseudo-scalar (left figure) and 
      vector meson masses (right figure).
   }
   \label{fig:meson:chiral_fit}
   \vspace{-5mm}  
\end{center}
\end{figure}

%// meson masses --------------------------------------------------------------

\section{Meson masses and decay constant}

Meson masses are extracted from correlation functions 
measured with smeared source and local sink operators.
Figure~\ref{fig:meson:chiral_fit} shows the chiral extrapolation for 
the pseudo-scalar and vector mesons made of degenerate valence quarks.
With our statistics, we do not observe any significant deviation
from simple linear fits
\bea
   M_{\rm PS}^2
   & = & 
   b_{\rm PS}\, m_{\rm sea},
   \hspace{5mm}
   M_{\rm V}
   = 
   a_{\rm V} + b_{\rm V}\,M_{\rm PS}^2,
   \label{eqn:meson:chiral_fit}
\eea
which give $\chi^2/{\rm dof} \lesssim 1.0$.
Employing this linear fit, 
we obtain $a\!=\!0.1312(23)$~fm from the $\rho$ meson mass input.
This is consistent with the estimate from $r_0$ within 10\% accuracy.

Since we are exploring small sea quark mass regime with a fixed topology,
we have to be careful about FSEs.
These are however difficult to assess reliably 
without performing direct simulations with different lattice volumes.
Here we only note that an analytic formula for FSE \cite{FSE:wrapping} 
combined with an analysis for the fixed topology \cite{fixed_topology} 
suggests that FSEs are not large ($\approx 1$\,--\,2\%) 
even at our lightest sea quark mass.

%// chiral symmetry breaking ------------------------------------------

\FIGURE{
   \centering
   % \begin{figure}[h!]
   % \begin{center}
   \includegraphics[angle=0,width=0.45\linewidth,clip]{AWIrat_vs_m.eps}
   \vspace{-3mm}
   \caption{
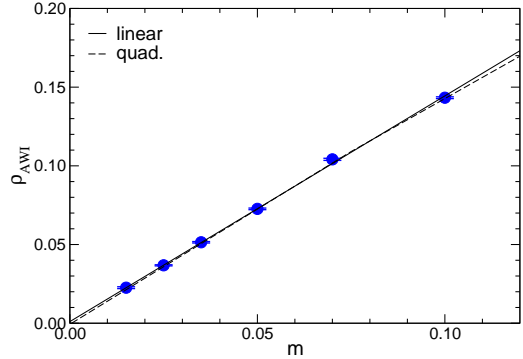
      Chiral extrapolation of ratio $\rho_{\rm AWI}$ 
      defined in Eq.~(\ref{eqn:meson:AWI_ratio}).
   }
   \label{fig:meson:AWI_ratio}
   % \end{center}
   % \end{figure}
}

In order to check the size of the chiral symmetry breaking, we calculate
the following ratio motivated from the axial ward identity
\bea
   \rho_{\rm AWI}
   & = &
   \frac{\langle \nabla_4 A_4 \, P^{\dagger} \rangle}
        {\langle P \, P^{\dagger} \rangle},
   \label{eqn:meson:AWI_ratio}
\eea   
where $P$ and $A_\mu$ are the pseudo-scalar and axial current operators.
Fig.~\ref{fig:meson:AWI_ratio} shows the chiral extrapolation of 
$\rho_{\rm AWI}$ with the following linear and quadratic forms:
\bea 
   \rho_{\rm AWI}
   = 
   a_{\rm AWI} + b_{\rm AWI}\,m_{\rm sea} 
               \left( + c_{\rm AWI}\,m_{\rm sea}^2 \right).
   &&
   \label{eqn:meson:AWI_ratio:chiral_fit}
\eea
The intercept $a_{\rm AWI}$ gives a measure of the chiral symmetry breaking.
We confirm that results from linear (0.0010(7)) and quadratic fits (-0.0007(9))
are consistent with zero.

The slope $b_{\rm AWI}$ provides a non-perturbative estimate of 
the renormalization constant $Z_A\!=\!2/b_{\rm AWI}$. 
We obtain 1.394(20)(81), where the first error is statistical,
and difference in $Z_A$ between the linear and quadratic extrapolations 
is added as a systematic error.

%// decay constant ----------------------------------------------------

\FIGURE{
   \centering
   % \begin{figure}[h!]
   % \begin{center}
   \includegraphics[angle=0,width=0.45\linewidth,clip]{fPS_vs_mPS2.r0.eps}
   \vspace{-3mm}
   \caption{
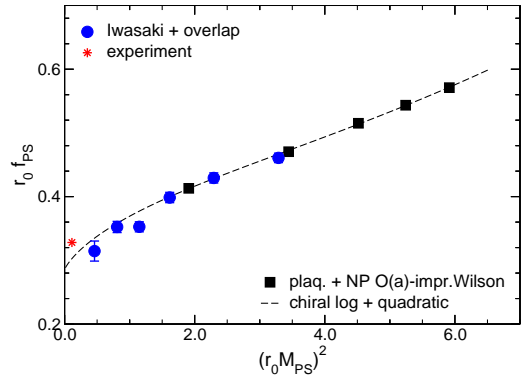
      Chiral behavior of $f_{\rm PS}$.
      % Vertical and horizontal axes are normalized by $r_0$.
      Squares are previous results with $O(a)$-improved Wilson quarks 
      \cite{Spectrum:Nf2QCD:JLQCD}. 
   }
   \label{fig:meson:f_pi}
   % \end{center}
   % \end{figure}
}

In Fig.~\ref{fig:meson:f_pi}, 
we plot our results for the pseudo-scalar decay constant 
together with our previous estimates 
with the $O(a)$-improved Wilson quarks 
\cite{Spectrum:Nf2QCD:JLQCD}.
All results are renormalized non-perturbatively 
by $Z_A=2/b_{\rm AWI}$ or $Z_A$ in Ref.\cite{NPZA}.
At heavy quark masses, we observe a reasonable consistency between 
our new and previous results with different discretizations.
This suggests that these data are close to their continuum limit
and have small scaling violations.

At small sea quark masses, however, 
its quark mass dependence is not smooth probably due to the limited statistics.
As a result, 
chiral extrapolation with the chiral logarithmic term is very unstable.
At this moment, it is difficult to make a definite conclusion 
on the consistency with chiral perturbation theory.

%// Summary -------------------------------------------------------------------

\section{Summary}

In this article, we report on the JLQCD's new project of large-scale simulations
with dynamical overlap quarks.
With careful choice of the lattice action and implementation of various 
algorithmic techniques, we demonstrated that it is feasible to accumulate
high statistics on reasonably fine ($a\!\approx\!0.125$~fm) and 
large lattices ($L\!\approx\!2$~fm).

Our simulations have just started. We are currently accumulating 
$O(10,000)$ trajectories for a precise determination of 
meson masses and decay constants, 
with which we are going to investigate
the consistency with chiral perturbation theory.

One of the most important subjects in the near future is 
the systematic effects due to the fixed topology. 
Simulations with $Q\!\ne\!0$ are underway to check $Q$ dependence 
of physical observables.
We are also planning to extend our simulations to three-flavor QCD.
Pushing simulations toward larger volumes is also interesting future direction
to explore the baryon sector as well as to check FSEs due to the small 
sea quark masses and the fixed topology.

%// Acknowledgments -----------------------------------------------------------

\vspace{5mm}

Numerical simulations are performed on Hitachi SR11000 and 
IBM System Blue Gene Solution 
at High Energy Accelerator Research Organization (KEK) 
under a support of its Large Scale Simulation Program (No.~06-13).
This work is supported in part by the Grant-in-Aid of the
Ministry of Education
(No.~13135204, 13135213, 13135216, 15540251, 16540228, 16740147, 16740156, 
     17340066, 17540259, 17740171, 18034011, 18104005, 18340075, 18740167).


\begin{thebibliography}{99}

\bibitem{Spectrum:Nf2QCD:JLQCD}
S.~Aoki {\it et al.} (JLQCD Collaboration),
Phys. Rev. D68 (2003) 054502.
% \vspace{-1mm}

\bibitem{Spectrum:Nf3QCD:CP-PACS+JLQCD}
T.~Ishikawa {\it et al.} (CP-PACS and JLQCD Collaborations),
PoS (LAT2006) 181.
% \vspace{-1mm}

% \bibitem{overlap}

% \bibitem{smearing:D_ov}

% \bibitem{iwasaki}

% \bibitem{mobility_edge}
% M.~Golterman, Y.~Shamir and B.~Svetitsky,
% Phys. Rev. D72 (2005) 034501.
% \vspace{-1mm}

\bibitem{lat06:JLQCD:yamada}
N.~Yamada {\it et al.} (JLQCD Collaboration), 
PoS (LAT2006) 060.
% \vspace{-1mm}

\bibitem{reflect}
Z.~Fodor, S.D.~Katz and K.K.~Szabo,
JHEP 0408 (2004) 003.
% \vspace{-1mm}

\bibitem{extra-Wilson:else}
P.M.~Vranas,
arXiv:hep-lat/0001006; hep-lat/0606014.
% \vspace{-1mm}

\bibitem{extra-Wilson:JLQCD}
H.~Fukaya {\it et al.} (JLQCD Collaboration),
arXiv:hep-lat/0607020.
% \vspace{-1mm}

\bibitem{lat06:JLQCD:hashimoto}
S.~Hashimoto {\it et al.} (JLQCD Collaboration), 
PoS (LAT2006) 052.
% \vspace{-1mm}

\bibitem{lat06:JLQCD:fukaya}
H.~Fukaya {\it et al.} (JLQCD Collaboration), 
PoS (LAT2006) 050.
% \vspace{-1mm}

\bibitem{multi_shift_solver}
A.~Frommer {\it et al.},
Int. J. Mod. Phys. C6 (1995) 627.
% \vspace{-1mm}

\bibitem{relaxed_solver}
N.~Cundy {\it et al.},
Comput. Phys. Commun. 165 (2005) 221.
% \vspace{-1mm}

\bibitem{mass_precond}
M.~Hasenbusch, 
Phys. Lett. B519 (2001) 177.
% \vspace{-1mm}

\bibitem{mtsMD}
J.C.~Sexton and D.H.~Weingarten,
Nucl. Phys. B380 (1992) 665.
% \vspace{-1mm}

\bibitem{mass_precond+mtsMD}
M.J.~Peardon and J.~Sexton,
Nucl. Phys. (Proc.Suppl.) 119 (2003) 985;
A.~Ali Khan {\it et al.} (QCDSF Collaboration),
Phys. Lett. B564 (2003) 235;
C.~Urbach, K.~Jansen, A.~Shindler and U.~Wenger,
Comput. Phys. Commun. 174 (2006) 87.
% \vspace{-1mm}

\bibitem{lat06:JLQCD:matsufuru}
H.~Matsufuru {\it et al.}, (JLQCD Collaboration), 
PoS (LAT2006) 031.
% \vspace{-1mm}

\bibitem{5D_solver}
A.~Bori\c{c}i, arXiv:hep-lat/9912040; hep-lat/0402035; 
R.G.~Edwards {\it et al.},
PoS (LAT2005) 146.
% \vspace{-1mm}

% \bibitem{smearing:pot}
% 

\bibitem{r0}
R.~Sommer, 
Nucl. Phys. B411 (1994) 839.
% \vspace{-1mm}

% \bibitem{r0:Nf0QCD:Takeda}

% \bibitem{r0:Nf2QCD:CP-PACS}

% \bibitem{unphys_phase_trans}

\bibitem{FSE:wrapping}
F.C.~Hansen and H.~Leutwyler,
Nucl. Phys. B350 (1991) 201.
% \vspace{-1mm}

\bibitem{fixed_topology}
R.~Brower, S.~Chandrasekharan, J.W.~Negele and U.-J.~Wiese,
Phys. Lett. B560 (2003) 64.
% \vspace{-1mm}

\bibitem{NPZA}
M.~Della Morte {\it et al.} (ALPHA Collaboration),
JHEP 0507 (2005) 007.

\end{thebibliography}
\end{document}